\documentclass[11pt,twoside,onecolumn]{article}
\usepackage[]{latexsym}
\usepackage{epsfig}
\usepackage{amsmath,amssymb}
\newcommand{\be}{\begin{eqnarray}}
\newcommand{\ee}{\end{eqnarray}}

\title{\bf Quantum Gravity Effects in Black Holes at the LHC}
\author{
G.L.~Alberghi$^{a,b,c}$\thanks{alberghi@bo.infn.it},
$\ $
R.~Casadio$^{a,b}$\thanks{casadio@bo.infn.it}
$\ $
and
A.~Tronconi$^{a,b}$\thanks{tronconi@bo.infn.it}
\\
\null
\\
$^a${Dipartimento di Fisica, Universit\`a di Bologna}
\\
{\em via Irnerio~46, 40126 Bologna Italy}
\\
$^b${Istituto Nazionale di Fisica Nucleare, Sezione di Bologna}
\\
$^c${Dipartimento di Astronomia, Universit\`a di Bologna}
\\
}
\begin{document}
\maketitle
\begin{abstract}
We study possible back-reaction and quantum gravity effects in
the evaporation of black~holes which could be produced at the LHC through a
modification of the Hawking emission.
The corrections are phenomenologically taken into account by employing a modified
relation between the black hole mass and temperature.
The usual assumption that black holes explode around $1\,$TeV is also released,
and the evaporation process is extended to (possibly much) smaller final masses.
We show that these effects could be observable for black holes produced
with a relatively large mass and should therefore
be taken into account when simulating micro-black hole events for the
experiments planned at the LHC.
\end{abstract}
\setcounter{page}{1}
\section{Introduction}
\setcounter{equation}{0}
One of the most exciting feature of models with large extra
dimensions~\cite{arkani,RS} is the fact that  the fundamental scale of
gravity $M_{\rm G}$ could be as low as the electro-weak scale
($M_{\rm G}\simeq 1$~TeV) and micro-black holes~\footnote[1]{Black holes
have been studied in both compact~\cite{argyres}
and infinite warped~\cite{Anchordoqui,chamblin} extra~dimensions
(see also Ref.~\cite{cavaglia}).}
may therefore be produced in our
accelerators~\cite{banks,Giddings3,argyres,ehm,dimopoulos}.
Once a black hole has formed (and after possible transients)
the Hawking radiation~\cite{hawking} is expected to set off.
The most common description of this effect is based on the
canonical~\footnote[2]{From a theoretical point of view, one should employ
the more consistent microcanonical description of black hole
evaporation~\cite{mfd} which was first applied in the context of large extra
dimensions in Refs.~\cite{bc,hossenfelder1,rizzo}.
It was then shown that actual life-times can vary greatly depending
on the model details~\cite{rizzo,bcLHC}.}
Planckian distribution for the emitted particles and on the consequence that
the life-time of  micro-black holes is so short that the decay can be viewed
as sudden~\cite{dimopoulos}.
This standard picture has already been implemented in several
numerical codes~\cite{charybdis,trunoir,ahn,catfish} which let the black
hole decay down to an arbitrary  mass $ M_{\rm f}$ of order
$M_{\rm G}$ via the Hawking law and then explode into a small
number of decay products.
By running these numerical codes, one mostly aims at extracting
information that will allow us to identify micro-black hole events in the
planned experiments based,
for example, on some experimental features of the decay
products~\cite{Hewett:2005iw} or on some particular signatures
of the decay~\cite{Humanic:2006xg}.
\par
We would like to stress here that the issue of the end of the black hole
evaporation remains an open question
(see, {\em e.g.},
Refs.~\cite{Landsberg:2006mm,Harris:2004xt,Casanova:2005id})
because we do not yet have a reliable theory of quantum gravity and,
most of all, no experimental data from the quantum gravity regime.
The singular behavior of the Hawking temperature as the black hole mass
decreases to zero can in fact be simply viewed as a sign of the lack of predictability
of perturbative approaches.
Therefore, although the very detection of these objects would already be
evidence that $M_{\rm G}$ is not the Planck energy (about $10^{19}\,$GeV) and
that extra dimensions may indeed exist, observing the late stages of black hole
evaporation (when the black hole mass $M\sim M_{\rm G}$) could provide
us with the kind of data we need to finally build the theory of
quantum gravity.
On a purely experimental side,
this is also an important issue, since it corresponds to determine
whether some features usually associated with the black hole decay
are sensitive to the assumptions made about the late stages of the
evaporation.
This is equivalent to asking whether deviations from the
Hawking law, induced by an underlying and still unknown theory of
quantum gravity, can actually be detected.
In some sense, this is analogous to the problem of trans-Planckian
modes in cosmology, where one can ask whether the still unknown
physics at Planck scales, which is certainly important in the early stages
of the Universe, can affect the CMB spectrum we observe now.
\par
As we mentioned above, the ignorance of the late stages of the black hole
evaporation is bypassed in the numerical codes by letting the micro-black hole
decay into a few standard model particles as an arbitrary lower mass
is reached (the possibility of ending the evaporation by leaving a
stable remnant has also been recently
considered~\cite{Koch:2005ks,Hossenfelder:2005bd}).
The purpose of this report is precisely to study if modifications of this standard
description of the evaporation which could occur as the black hole mass becomes
smaller than some arbitrary scale ($M\lesssim M_{\rm G}$) can produce
experimentally detectable features,
although we shall not consider a specific experimental setup and leave for future
developments the task of including the detector's sensitivity and geometry.
From another perspective, one can also view this preliminary
analysis as an estimate of the systematic errors to be associated with
the results already appeared on this subject that usually rely on
the standard picture we described above.
\par
For this purpose, we have developed several modifications to the Monte Carlo
code CHARYBDIS~\cite{charybdis} which we shall describe in Section~\ref{TTGM}
(see also Ref.~\cite{LHCb}).
We shall then report some interesting and perhaps unexpected (preliminary)
results from a few test runs in Section~\ref{sim}.
We shall use units with $c=\hbar=1$ and the Boltzmann constant $k_{\rm B}=1$.
\section{Black hole evaporation}
\setcounter{equation}{0}
No-hair theorems of General Relativity guarantee that a black hole is
characterized by its mass, charges and angular momentum only.
The one parameter characterizing an uncharged, non-rotating black hole 
is thus its mass $M$.
On considering solutions to the Einstein equations (or applying Gauss' theorem)
in $4+d$ dimensions, one can then derive the following relation between the mass
and the horizon radius,
\be
R_{\rm H}=\frac{1}{\sqrt{\pi}\,M_{\rm G}}\,
\left(\frac{M}{M_{\rm G}}\right)^{\frac{1}{d+1}}
\left(\frac{8\,\Gamma\left(\frac{d+3}{2}\right)}{d+2}
\right)^{\frac{1}{d+1}}
\ ,
\ee
where $\Gamma$ is the usual Gamma function,
and the temperature associated with the horizon
is given by
\be
T_{\rm H} =
\frac{d+1}{4\,\pi\,R_{\rm H}}
\ .
\label{TH}
\ee
Once formed, the black hole begins to evolve.
In the standard picture the evaporation process can be divided
into three characteristic stages~\cite{Giddings3}:
\begin{enumerate}
\item {\sc Balding phase:}
the black hole radiates away the multipole moments
it has inherited from the initial configuration,
and settles down in a hairless state.
A certain fraction of the initial mass will also be
lost in gravitational radiation.
\item {\sc Evaporation phase:}
it starts with a spin down phase in which the Hawking
radiation~\cite{hawking} carries away the angular momentum, after
which it proceeds with the emission of thermally distributed
quanta until the black hole reaches the Planck mass (replaced by
the fundamental scale $M_{\rm G}$ in the models we are considering
here).
The radiation spectrum contains all the Standard Model particles,
which are emitted on our brane, as well as gravitons, which are
also emitted into the extra~dimensions.
It is in fact expected that most of the initial energy is emitted
during this phase into Standard Model particles~\cite{ehm}
(although this conclusion is still being debated, see
{\em e.g.},~Ref.~\cite{cavaglia2}).
\item {\sc Planck phase:}
once the black hole has reached a mass close to the effective Planck
scale $M_{\rm G}$, it falls into the regime of quantum gravity and
predictions become increasingly difficult.
It is generally assumed that the black hole will either completely
decay into some last few Standard Model particles or a stable remnant
be left which carries away the remaining energy~\cite{Koch:2005ks}.
\end{enumerate}
\par
In our approach we will consider possible modifications to the
second and third phases.
On the one hand, we will allow a modified Hawking phase by
employing a different relation between the
horizon radius and the temperature.
The necessity of such a modification, suggested by the fact that the
Hawking temperature diverges as the mass of the black hole goes to zero,
has been discussed in several papers, {\em e.g.},
Refs.~\cite{modifications, nicolini}.
We will follow a phenomenological approach to this issue and model
a general class of modifications by making use of the form given in
Eq.~(\ref{Tmod}) below.
On the other hand, we will look at the possibility that the evaporation
may not end at the fundamental scale $M_{\rm G}$ ($\sim 1\,$TeV)
but proceeds further until a lower (arbitrary)
mass $M_{\rm f} $ has been reached.
\section{Quantum Gravity and Monte Carlo code}
\setcounter{equation}{0}
\label{TTGM}
Several Monte Carlo codes which simulate the production and decay
of micro-black holes are now available  (see, {\em
e.g.}~Refs.~\cite{charybdis,trunoir,ahn}).
We have found it convenient to implement our modifications into
CHARYBDIS~\cite{charybdis}.
\par
In order to let the black hole evaporate below
the electro-weak scale $M_{\rm EW}$, at which the Standard Model
particles acquire their mass, one needs to treat the mass of the
emitted particles properly.
In the original code, the phase space was taken to be that of massless
particles since the mass of the black hole $M\gtrsim M_{\rm f}$ and
$M_{\rm f}>M_G>M_{\rm EW}$.
However, we also want to consider the possibility that the black hole
decays to lower masses, that is $M_{\rm f}\ll M_G $. 
We therefore use the phase space measure in the Planckian number density
for the emitted particles given by
\be
N_m(\vec k)=\frac{d^3 k}{e^{\omega/T}\pm 1}
\ ,
\label{N_k}
\ee
where $T$ denotes the black hole temperature, $m$ the particle mass,
$\vec k$ the particle 3-momentum and $\omega=\sqrt{k^2+m^2}$.
Moreover, since $N_m$ depends on $m$ and not just on the
statistics, one can no more assume (as in CHARYBDIS)
a fixed ratio of production for
fermions versus bosons  but proper particle
multiplicity must be used when generating particle types randomly.
We used the multiplicities predicted by the Standard
Model~\cite{pdg} and modified the code accordingly.
\par
\begin{figure}[ht]
\centerline{
\raisebox{4cm}{$T$}
\epsfxsize=200pt\epsfbox{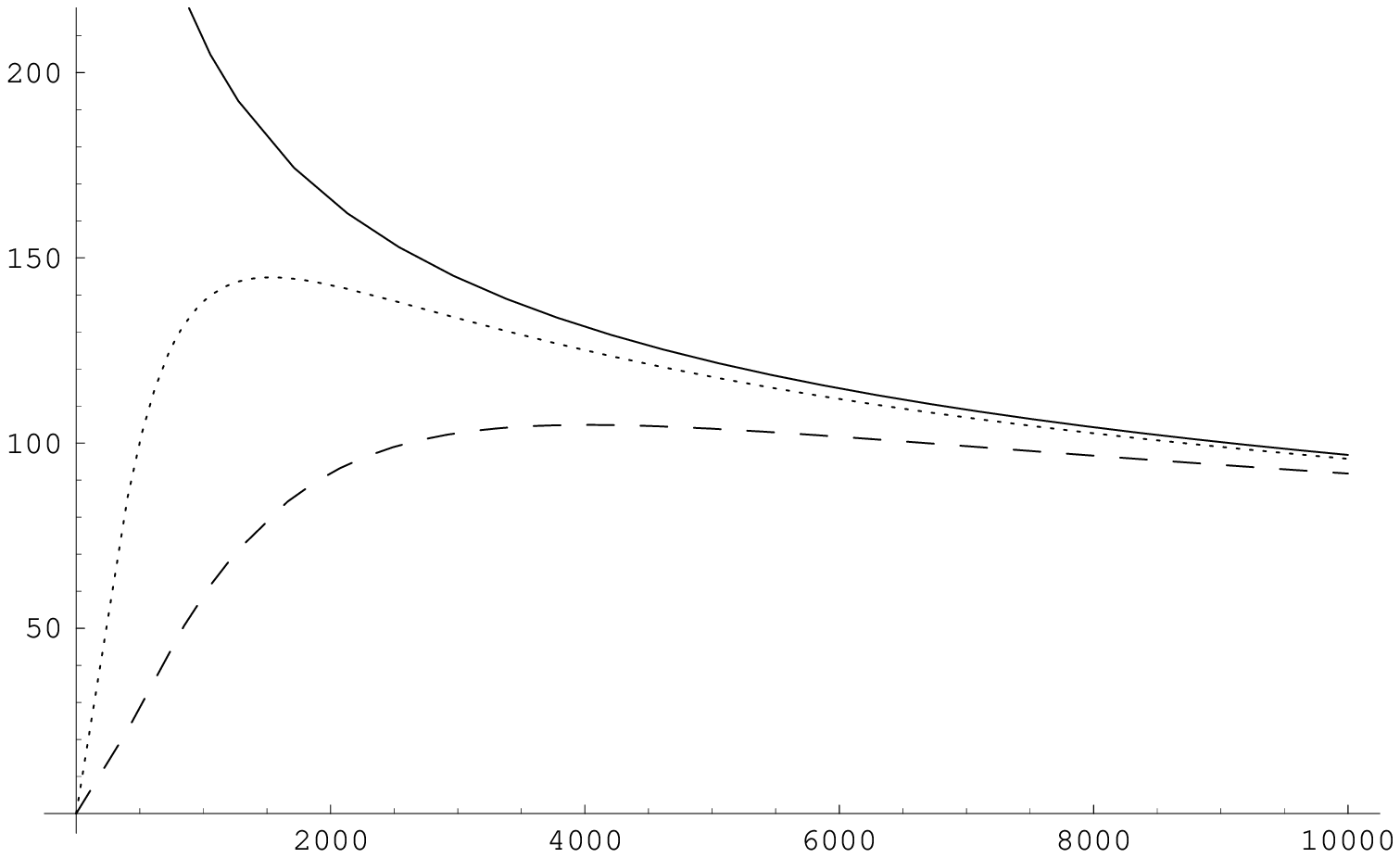}
\hspace{0.5cm}
\raisebox{4cm}{$T$}
\epsfxsize=200pt\epsfbox{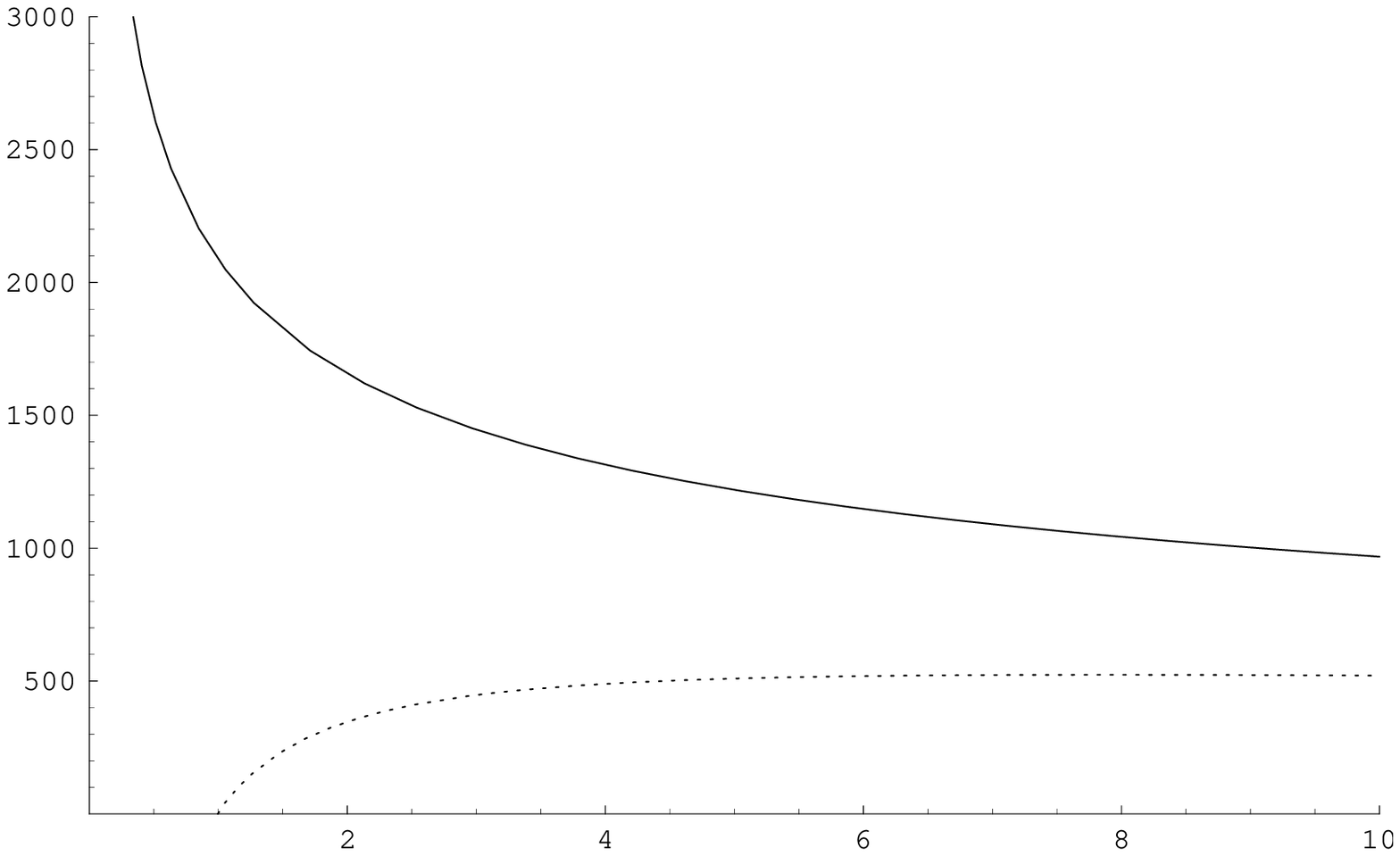}
}
\hspace{7cm}
$M$
\hspace{7cm}
$M$
\caption{Black hole temperature (in GeV) as a function of the black hole mass (in GeV).
Left panel:
comparison between the Hawking law $T_{\rm H}$ (solid line)
and $T_1$ from Eq.~(\ref{Tmod1})
for $\ell=10^{-3}\,$GeV$^{-1}$, $\alpha=5$, $n=5$ (dashed line)
and $\ell=10^{-3}\,$GeV$^{-1}$, $\alpha=5$, $n=1$ (dotted line).
Right panel:
$T_{\rm H}$  (solid line) compared to $T_2$
from Eq.~(\ref{Tmod2}) with
$\ell=1.14\cdot 10^{-4}\,$GeV$^{-1}$, $n_1=1$, $n_2=1$ (dotted line).
\label{T}
}
\end{figure}
In order to include possible quantum gravity effects, we employ
modified expressions for the temperature $T=T_i$ of the form
\be
T_i=F_i
\,T_{\rm H}
\label{Tmod}
\ ,
\ee
in which, in order to cover results in the
existing literature (for a partial list of approaches to the
problem, see Refs.~\cite{mfd,hossenfelder1,rizzo,bcLHC,nicolini})
we shall consider the functional forms
\be
F_1(\ell,\alpha,n)=
\frac{R_{\rm H}^n}{R_{\rm H}^n+\alpha\,\ell^n}
\label{Tmod1}
\ee
and
\be
F_{2}(\ell,n_1,n_2)=
\left[1-\left(\frac{\ell}{R_H}\right)^{n_1}
\right] ^{n_2}
\ ,
\label{Tmod2}
\ee
where $\ell$, $\alpha$, $n$, $n_1$, and $n_2$ are parameters
that can be adjusted (see below).
Note that $F_1 $ leads to a vanishing temperature for vanishing
black hole mass (horizon radius) whereas with $F_2$ the
temperature vanishes at finite $M$ when $R_{\rm H}=\ell$~\cite{nicolini}
(such a remnant was also considered in Ref.~\cite{Koch:2005ks}).
Specific examples which will be used throughout the paper
are shown in Fig.~\ref{T} together with the
standard Hawking law~(\ref{TH}).
\par
A list of some adjustable parameters in the code is given in Table~\ref{FreePar}.
The initial black hole mass $M_0$ can be either fixed (to within the maximum
centre mass energy expected at the LHC) or generated according to the
partonic cross section.
In the simulations we present here, we shall take a total centre mass energy
of $14\,$TeV and mostly focus on the the case of $M_0= 10\,$TeV (with
$M_0=2\,$TeV taken for comparison~\footnote{According
to the current understanding of black hole production, light black holes of mass
around $2\,$TeV are more likely to be produced at the LHC (see Ref.~\cite{dimopoulos})
but detailed features in their decay signal would also be less clearly identifiable
than in those of larger mass.}).
At the end point of the decay, the black hole
explodes in a selectable number $N_{\rm f}$ of fragments
when $M\lesssim M_{\rm f}$.
We shall use $N_{\rm f} = 2$ and set the total number of space-time
dimensions equal to 6.
We shall also set the grey-body factors equal to 1 for all kinds of particles.
This approximation is certainly restrictive and calls for a more refined analysis.
However, we note that in the standard picture the use of more
accurate grey-body
factors alters the final results only by a small amount and we thus feel
that our modified code can be regarded as a good tool to study qualitative
features related to the late stages of the evaporation.
A substantial modification of CHARYBDIS will likely be necessary if one wants to obtain more accurate predictions about, for example, the actual modification of the Hawking
emission.
\begin{table}[h]
\centerline{
\begin{tabular}{|c|c|c|}
\hline
$M_0$
&
Initial black hole mass
&
$1-14$\,TeV; random
\\
\hline
$M_{\rm f}$
&
Minimum black hole mass
&
$1-1000$\,GeV
\\
\hline
$N_{\rm f}$
&
number of final fragments
&
$2-6$
\\
\hline
$d$
&
number of extra dimensions
&
$1-5$
\\
\hline
$F_i$
&
modified temperature
&
$1$ (no modification), $F_1$, $F_2$
\\
\hline
\end{tabular}
}
\caption{Parameters that can be adjusted in our
Monte Carlo generator and their ranges.
For each modified temperature, one can also set the corresponding
parameters as described in the text.}
\label{FreePar}
\end{table}
\section{Simulation Results}
\setcounter{equation}{0}
\label{sim}
The standard CHARYBDIS generator simulates the evaporation of a
micro-black hole according to Hawking's law until the black
hole mass $M$ reaches the minimum value $M_{\rm f}$ which is assumed
equal to the fundamental scale of gravity $M_{\rm G}$ in
a given model of extra dimensions ({\em e.g.}, at $1\,$TeV).
The black hole subsequently decays into a few bodies ($N_{\rm f}=1,\ldots,6$)
simply according to phase space.
As described in the previous section, in our modified code,
we implemented ways to extend the evaporation to a mass $M_{\rm f}$
well below the fundamental scale ({\em e.g.}, down to $1\,$GeV).
\par
In the following, we shall compare the outputs of the standard and
modified generators from a phenomenological point of view.
In particular, we shall consider the output of the standard code
with $M_{\rm f}=1\,$TeV (and $M_{\rm f}=1\,$GeV for completeness)
as reference.
The results for the modified temperature $T_1$
were obtained with $M_{\rm f}=1\,$GeV and the parameters
$\ell=10^{-3}\,$GeV$^{-1}$, $\alpha=5$, $n=5$ (see Eq.~(\ref{Tmod1}))
and those for the temperature $T_2$
with $\ell=1.14\cdot 10^{-4}\,$GeV$^{-1}$, $n_1=1$, $n_2=1$
(see Eq.~(\ref{Tmod2})).
Note that the latter choice of parameters also corresponds to a final mass of
the black hole equal to $M_{\rm f}\simeq 1\,$GeV at which the temperature $T_2=0$.
These cases are the same that were shown previously in Fig.~\ref{T}.
\subsection{Primary Emission}
\begin{figure}[ht]
\centerline{
\epsfxsize=220pt\epsfbox{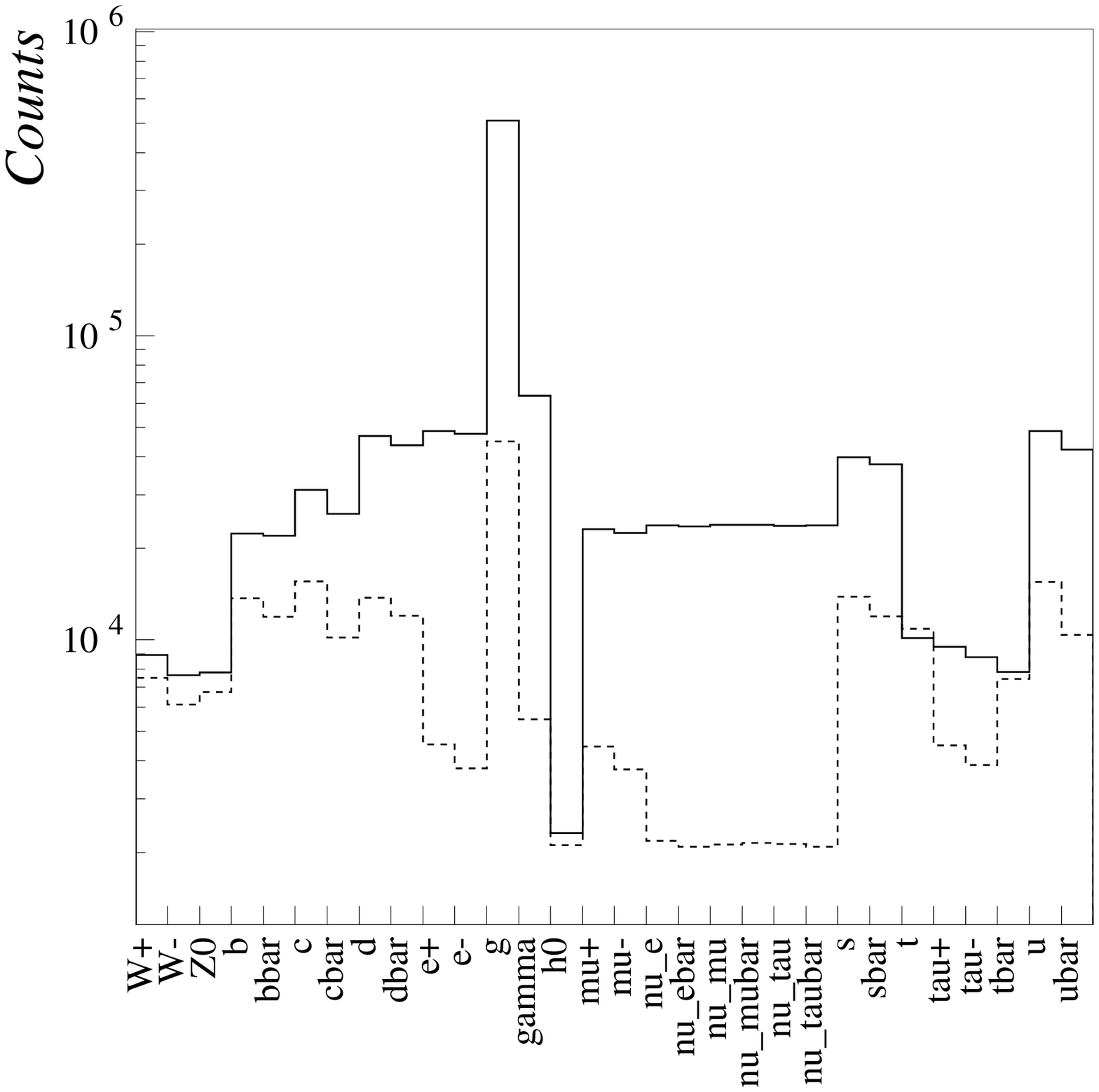}
\hspace{0.5cm}
\epsfxsize=220pt\epsfbox{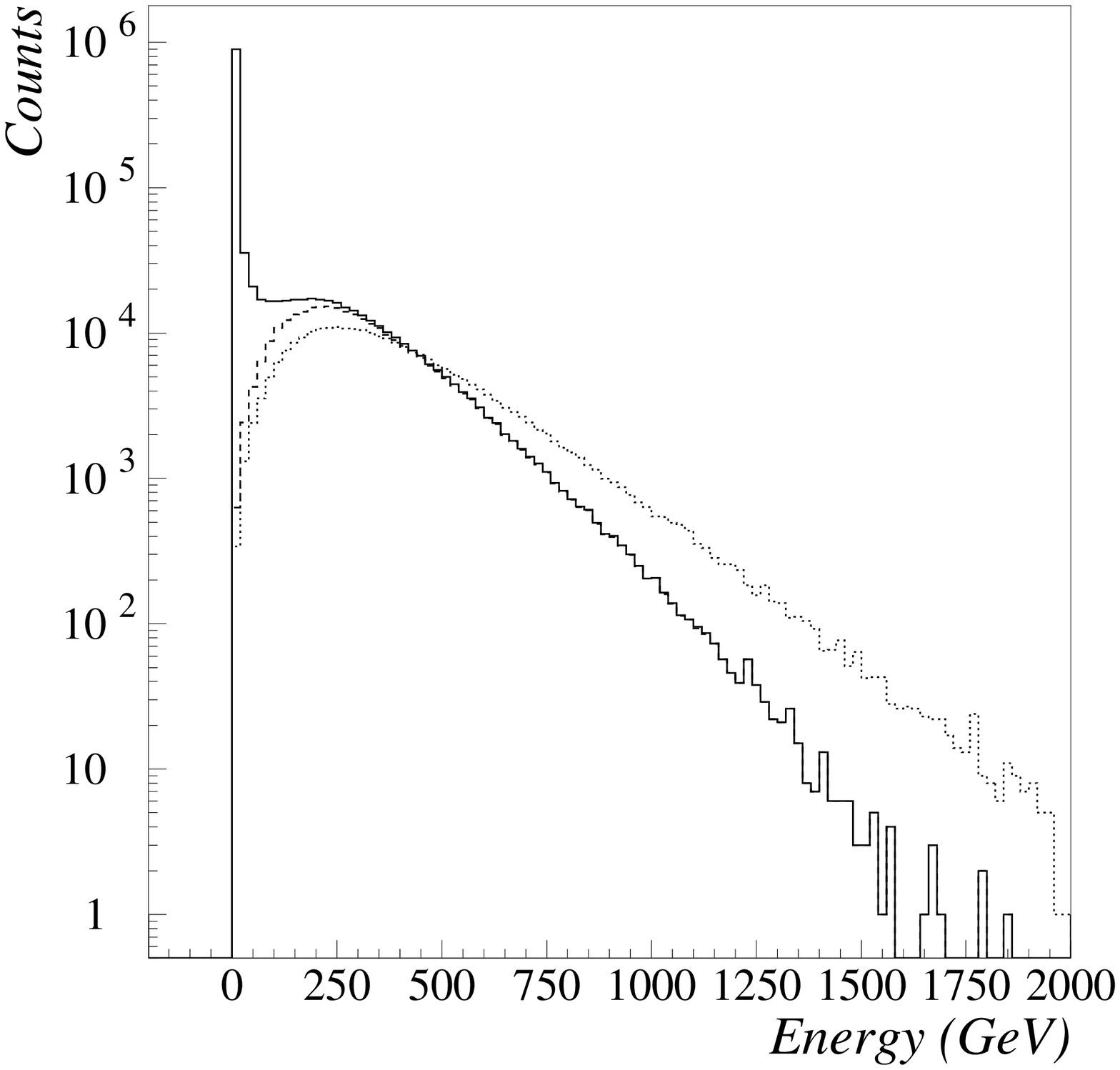}
}
\caption{Left panel: abundance of Standard Model particles produced
during the evaporation of $10\,$TeV black holes before hadronization,
for a run of 10000 events with the temperature $T_1$ (solid histogram) and with the
standard CHARYBDIS generator (dashed histogram).
The fundamental scale of gravity is set to $1\,$TeV, temperature
parameters are the same as in Fig.~\ref{T} and number of extra dimensions $d=2$.
Right panel: energy distribution of the emitted particles obtained
from $T_1$ (solid histogram), including only
particles emitted when the black hole has a mass exceeding $1\,$TeV
(dashed histogram), and with the standard CHARYBDIS generator
(dotted histogram).
\label{panam}
}
\end{figure}
\begin{figure}[ht]
\centering{
\includegraphics[width=0.45\textwidth]{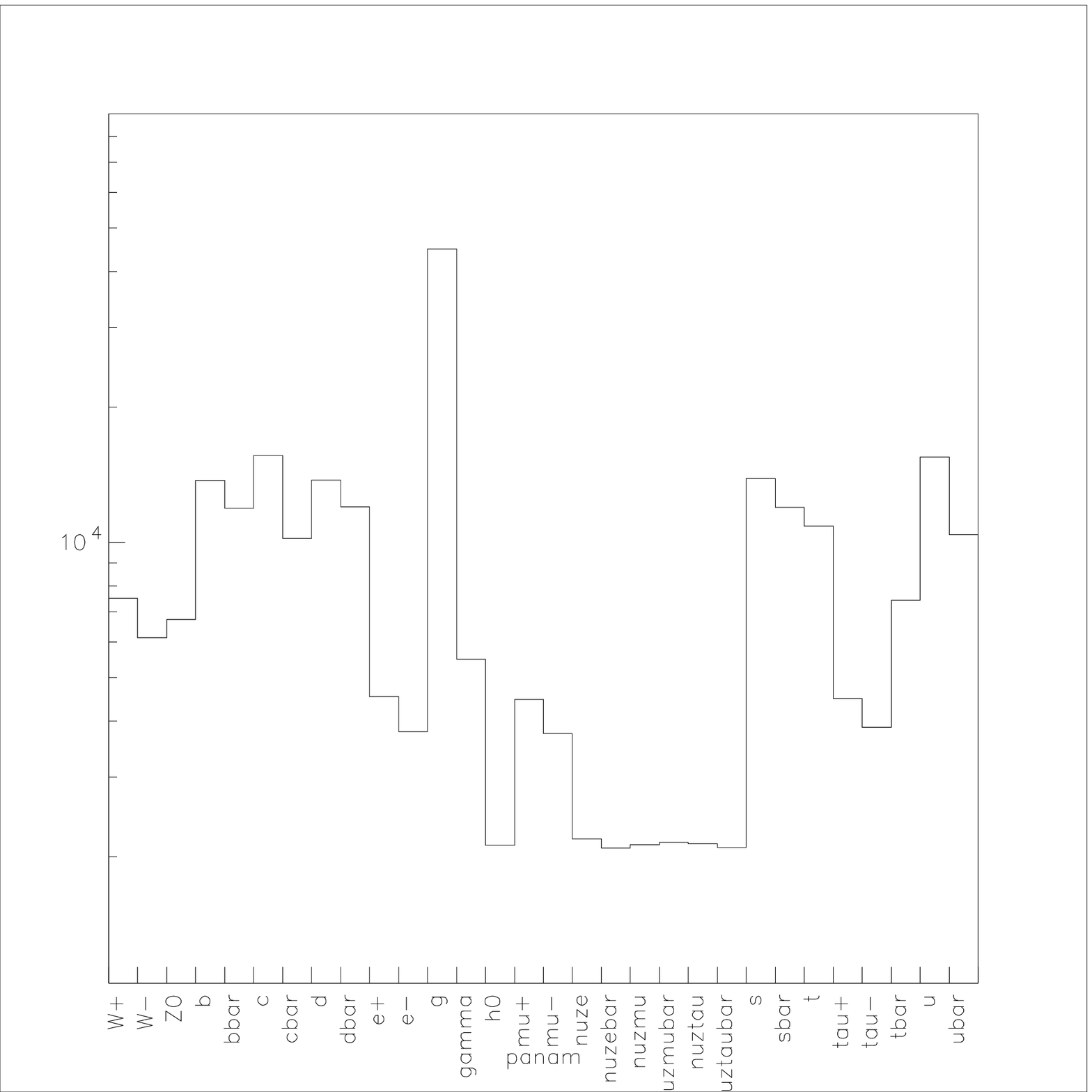}
\hspace{0.5cm}
\includegraphics[width=0.45\textwidth]{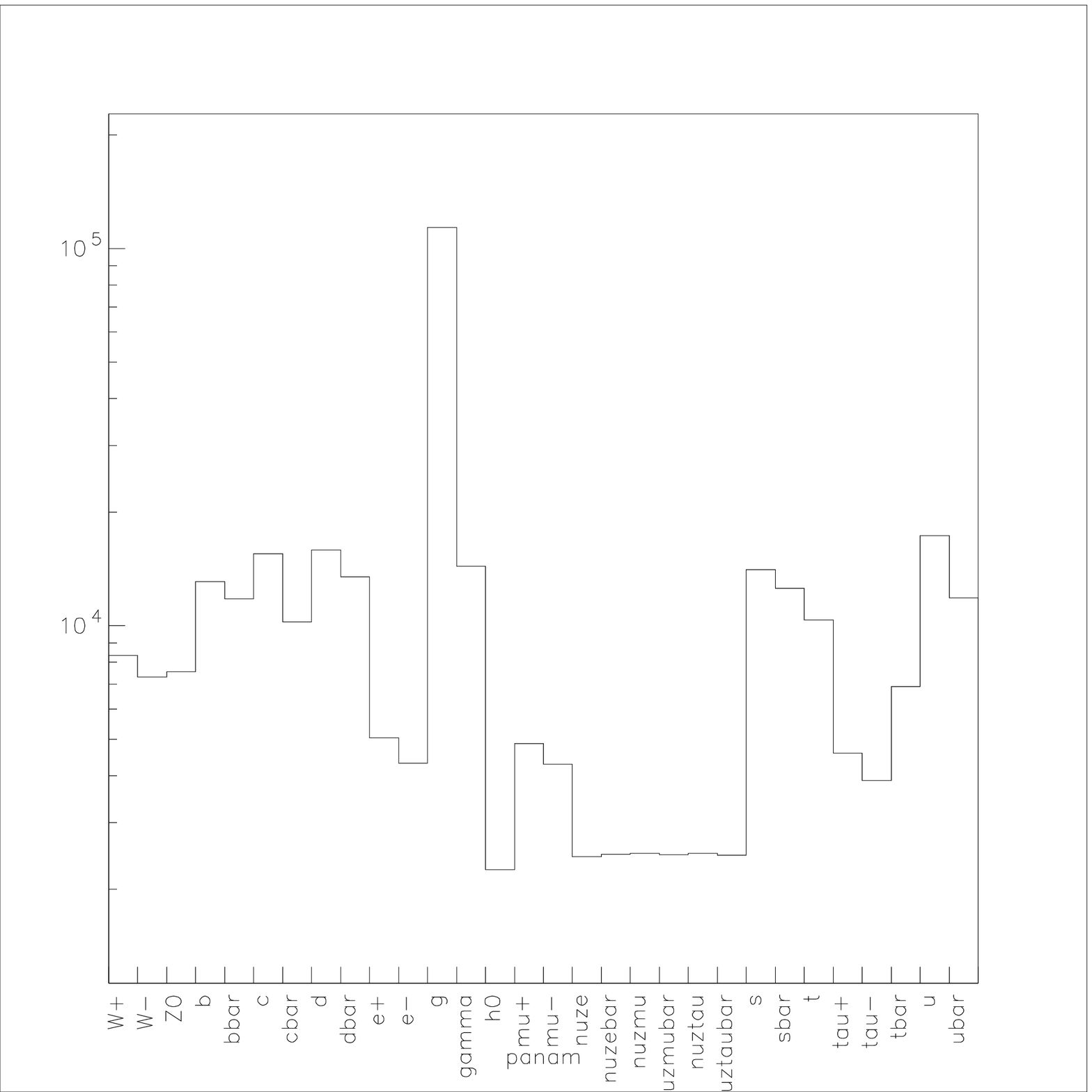}
} \vspace{0.5cm}
\\
\centering{
\includegraphics[width=0.45\textwidth]{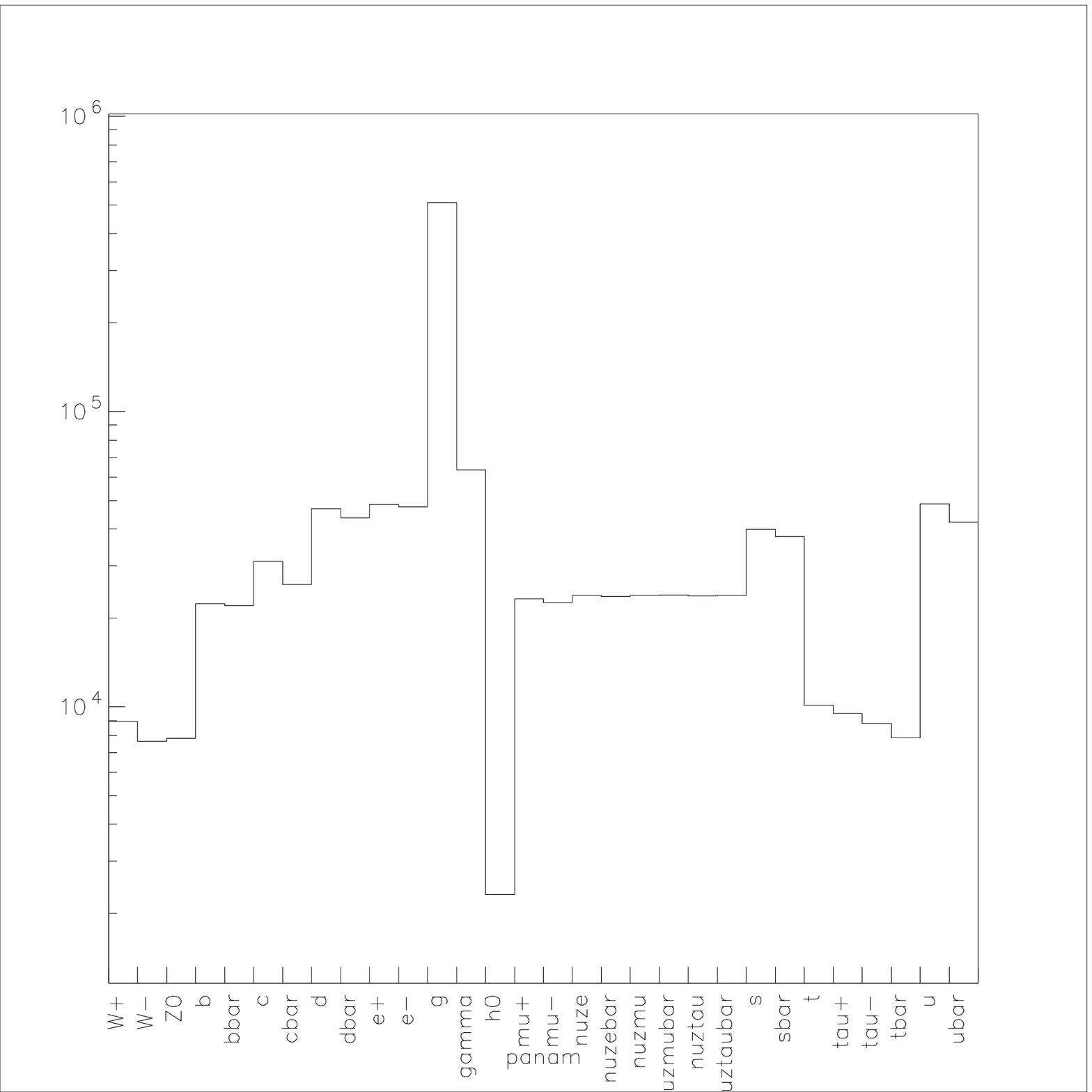} 
\hspace{0.5cm}
\includegraphics[width=0.45\textwidth]{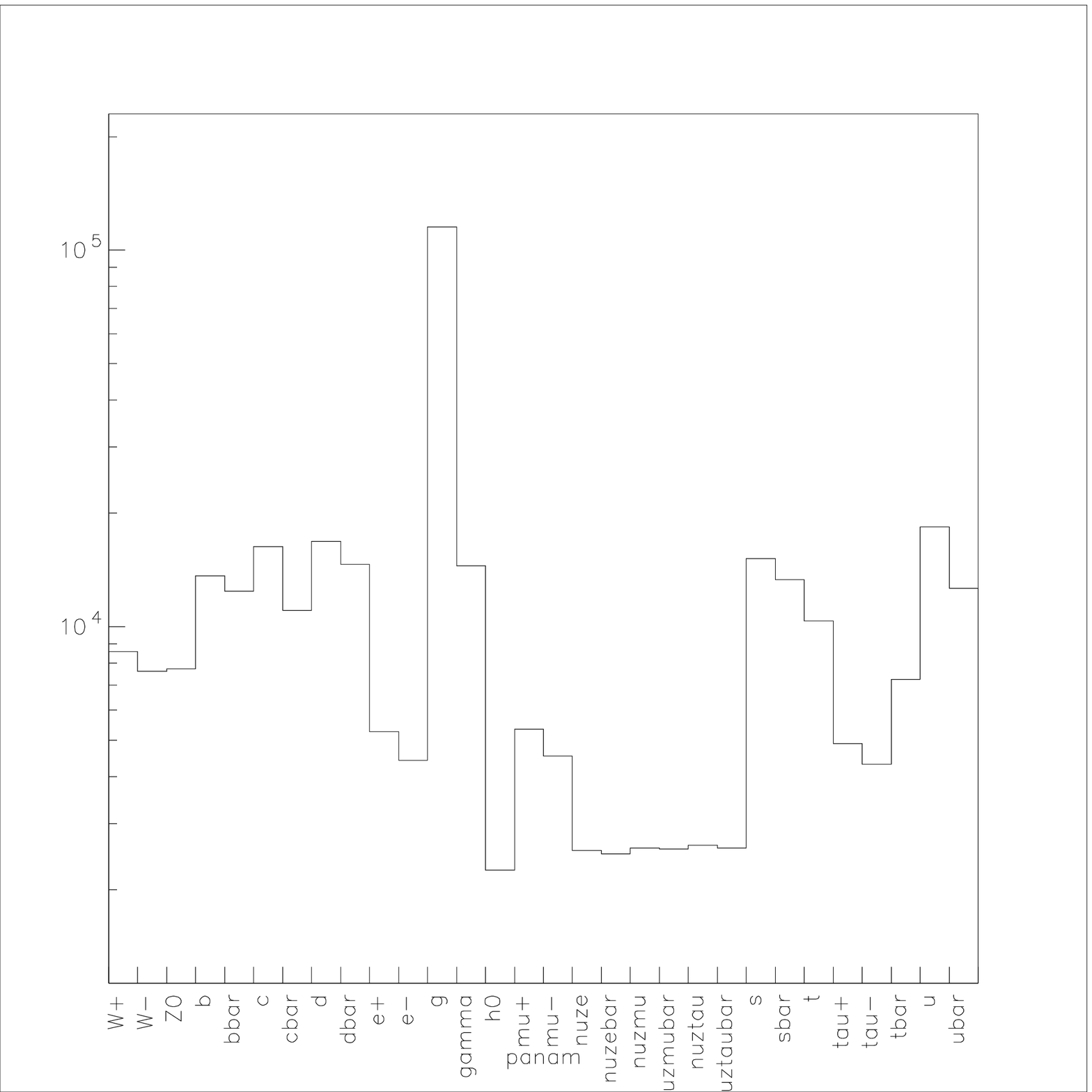}
}
\caption{Abundance of Standard Model particles produced during the
evaporation of $10\,$TeV black holes for a run of 10000 events before
hadronization.
Upper graphs:
standard CHARYBDIS generator with $M_{\rm f} =1\,$TeV
(left panel) and $M_{\rm f} =1\,$GeV (right panel).
Lower graphs:
modified code with the temperature $T_1$ (left panel)
and $T_2$ (right panel) and $M_{\rm f} =1\,$GeV.
\label{part50} }
\end{figure}
\begin{figure}[ht]
\centering{
\includegraphics[width=0.45\textwidth]{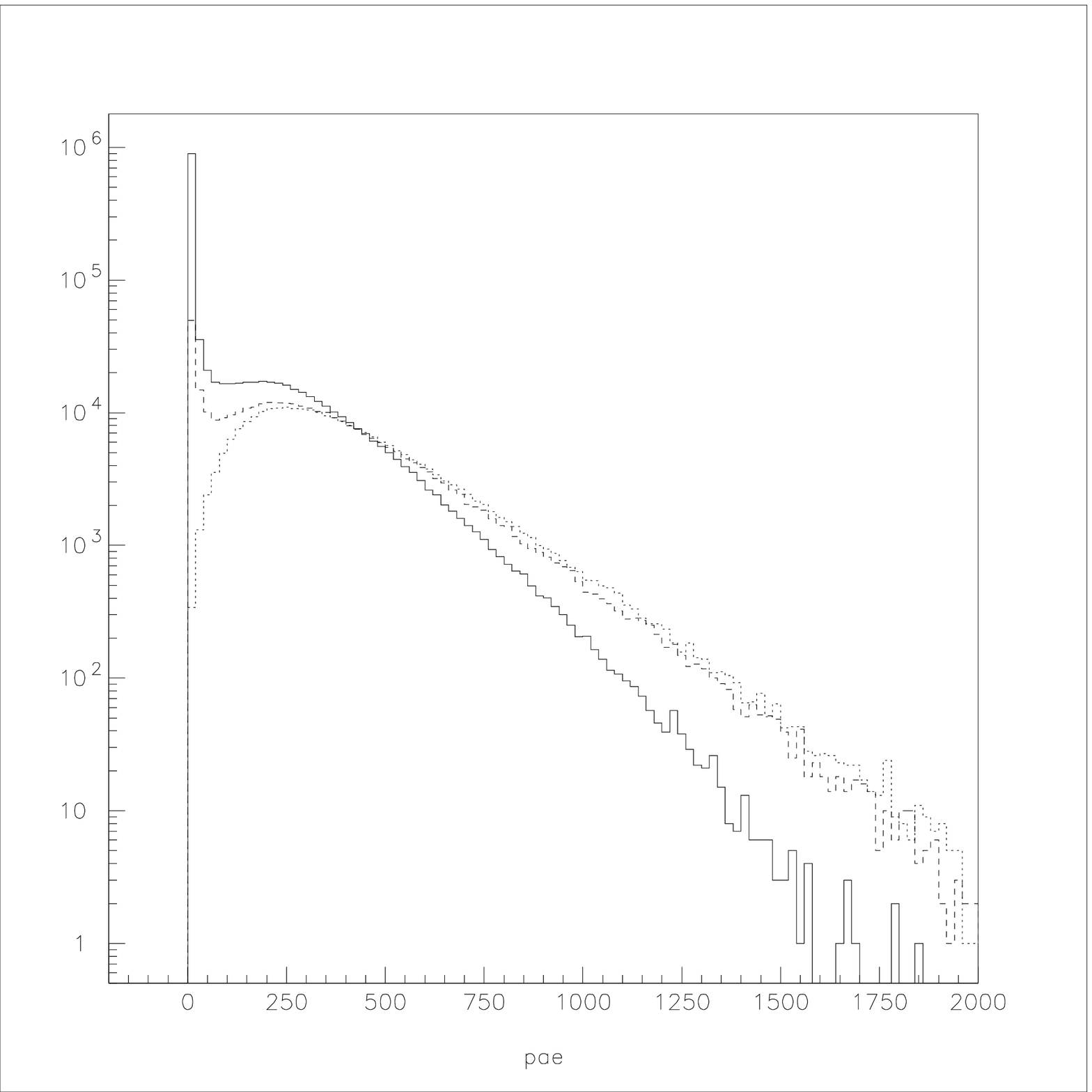}}
\caption{Energy spectrum of particles produced during the
evaporation of $10\,$TeV black holes for a run of 10000 events before
hadronization:
standard CHARYBDIS generator with $M_{\rm f} =1\,$TeV
(dotted line) and $M_{\rm f} =1\,$GeV (dashed line),
modified code with the temperature $T_1$ (solid line)
and $T_2$ (indistiguishable from the dashed).
\label{pae50} }
\end{figure}
Let us first examine the ``primary'' black hole emission, namely the
particles produced by direct black hole evaporation before parton
evolution and hadronization are taken into account.
As we mentioned before, we set the mass for the final decay of the black hole at
$M_{\rm f} = 1\,$TeV in the standard case and at $M_{\rm f}= 1\,$GeV
for the modified temperatures.
\par
A comparison of the relative abundance of the Standard Model particles
produced by the black hole evaporation with the Hawking temperature
and the modified temperature $T_1$ from Eq.~(\ref{Tmod1}) is shown
in the left panel of Fig.~\ref{panam}.
For the temperature $T_1$, this plot clearly shows a much larger multiplicity
of isotropically emitted light particles ({\em i.e.}, photons and neutrinos as well as
electrons and muons).
In the right panel of Fig.~\ref{panam}, we also show the energy distribution
of all the emitted particles.
The modified law dramatically changes the spectrum at low energy
(there are about three orders of magnitude more particles with energy
below $100\,$GeV than in the standard case),
leaving the spectrum at large energy moderately affected
(by just about one order of magnitude).
Note that the low energy distribution of particles emitted when the black hole
mass $M>1\,$TeV follows closely the standard curve but that there are
already missing particles in the high energy tail (above $1\,$TeV).
\par
For completeness, in Fig.~\ref{part50}, we show the particle abundance
obtained with the standard CHARYBDIS generator if we set $M_{\rm f}=1\,$GeV
and compare it with the results from the modified code for the two different
temperatures $T_1$ and $T_2$ and $M_{\rm f}=1\,$GeV.
It then appears that the temperature $T_1$ leads to a number of light
particles significantly larger than the other three cases.
The temperature $T_2$ instead produces about the same amount
of light particles as the standard Hawking temperature provided the evaporation is
extended down to $M_{\rm f}=1\,$GeV.
The corresponding energy distributions are given in Fig.~\ref{pae50}, from
which one concludes that the number of high energy particles produced with
$T_2$ is also closer to the standard case with $M_{\rm f}=1\,$GeV.
To conclude, if one just considers the above data, there is no visible difference
in the primary emission between $T_2$ and the Hawking temperature
provided both are extended to a minimum black hole mass of $1\,$GeV,
whereas the spectrum generated with $T_1$ still shows appreciable
differences.
\par
These behaviors can be understood in the following way:
since the temperature $T_1$ tends to zero for $M=0$, we do expect
a continuous emission of increasingly softer particles until the black hole
evaporates completely ({\em i.e.}, it reaches the minimum value
$M_{\rm f}=1\,$GeV).
Moreover, once the black hole mass has decreased below the
threshold for producing a given massive particle, such a particle
can no longer be emitted.
Hence one also expects that the production of the heavier particles
({\em i.e.},~massive gauge bosons and top quarks)
be scarcely affected, while the emission of soft low mass (or
massless) particles should be largely enhanced.
For the temperature $T_2$, the situation is similar except that
its functional form follows the Hawking law more closely
and the distribution of particles more closely resembles the
standard case provided one allows the black hole to evaporate
down to the same mass at which $T_2$ vanishes ($1\,$GeV
in our samples).
\subsection{Final Output}
\begin{figure}[ht]
\centering{
\includegraphics[width=0.45\textwidth]{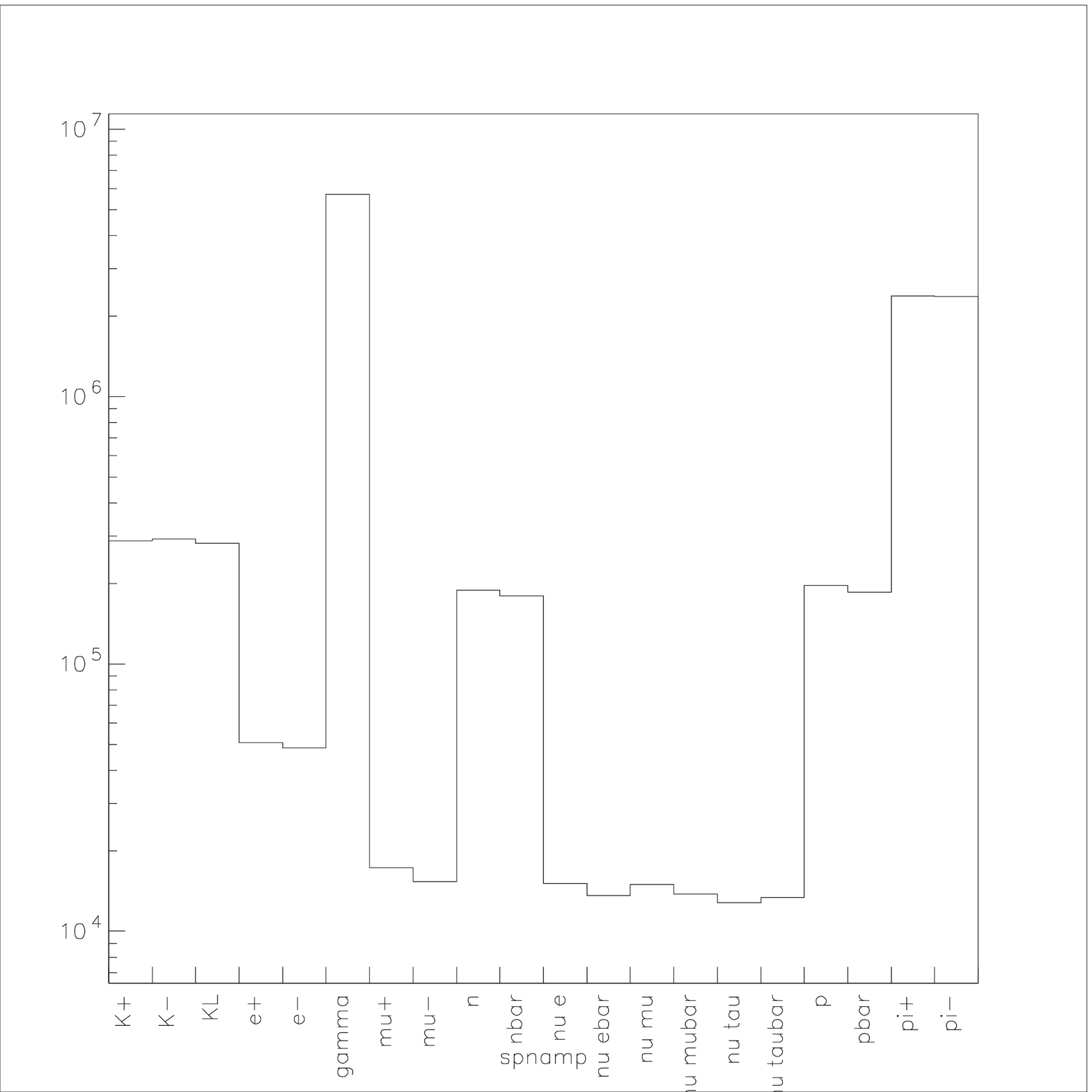}
\hspace{0.5cm}
\includegraphics[width=0.45\textwidth]{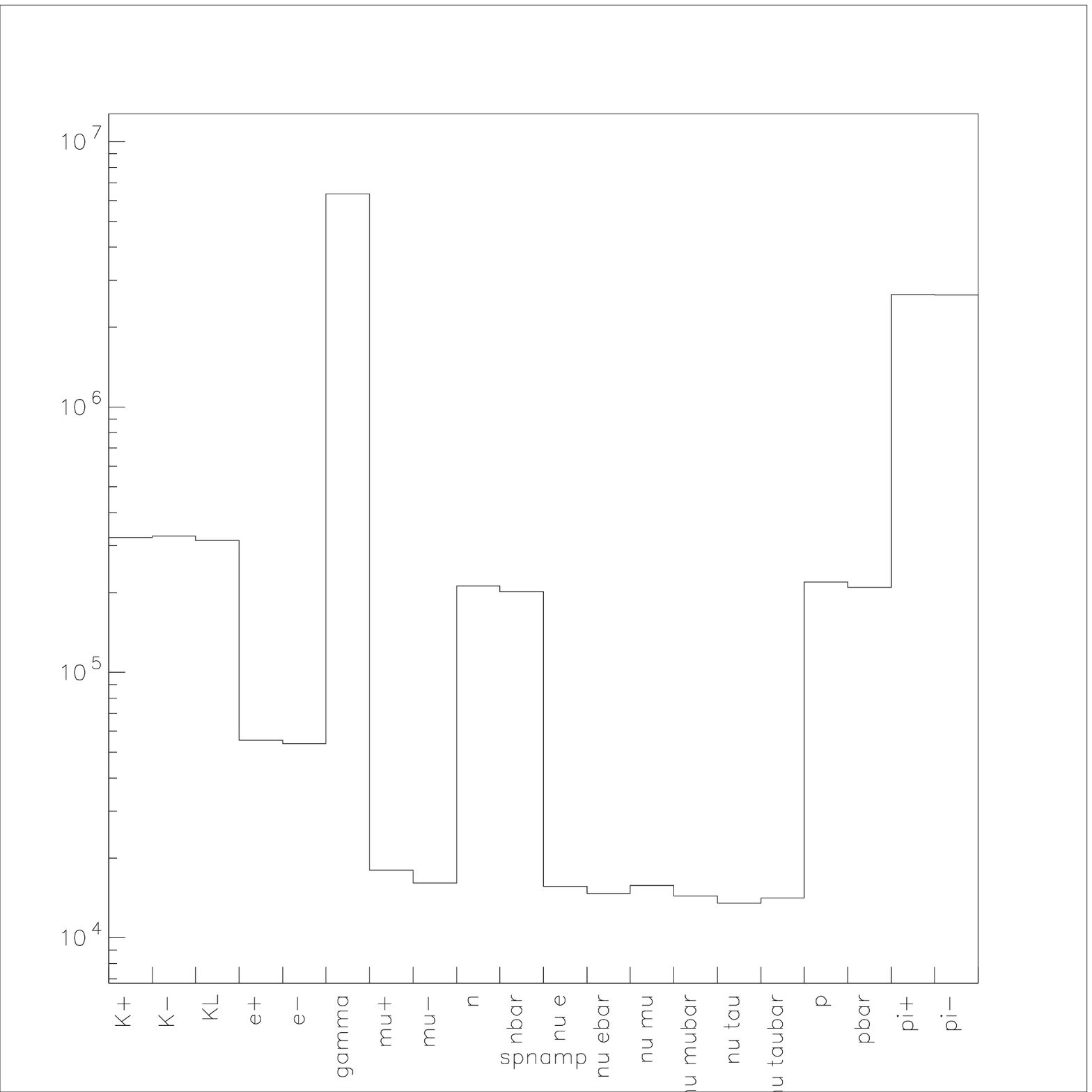}
} \vspace{0.5cm}
\\
\centering{
\includegraphics[width=0.45\textwidth]{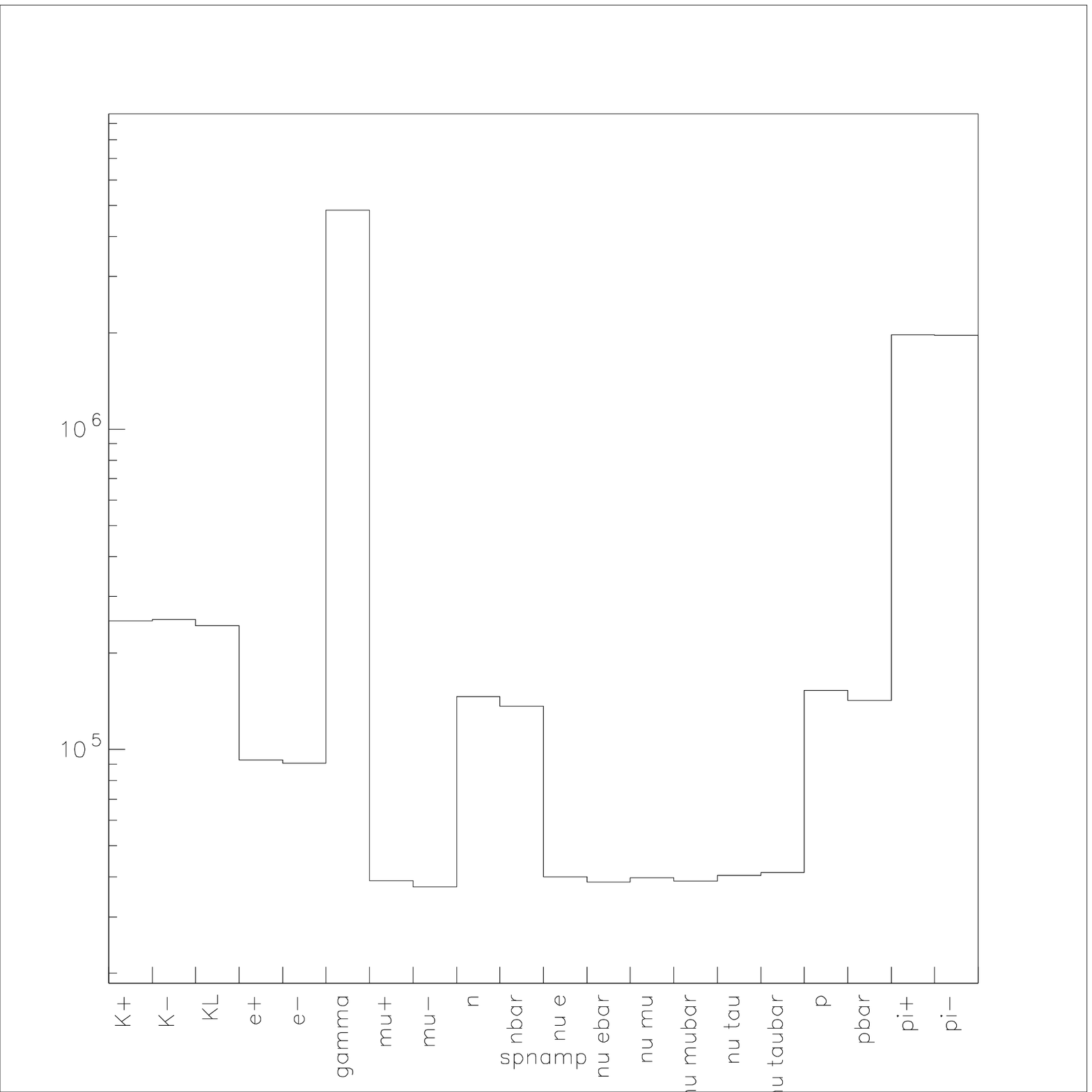}
\hspace{0.5cm}
\includegraphics[width=0.45\textwidth]{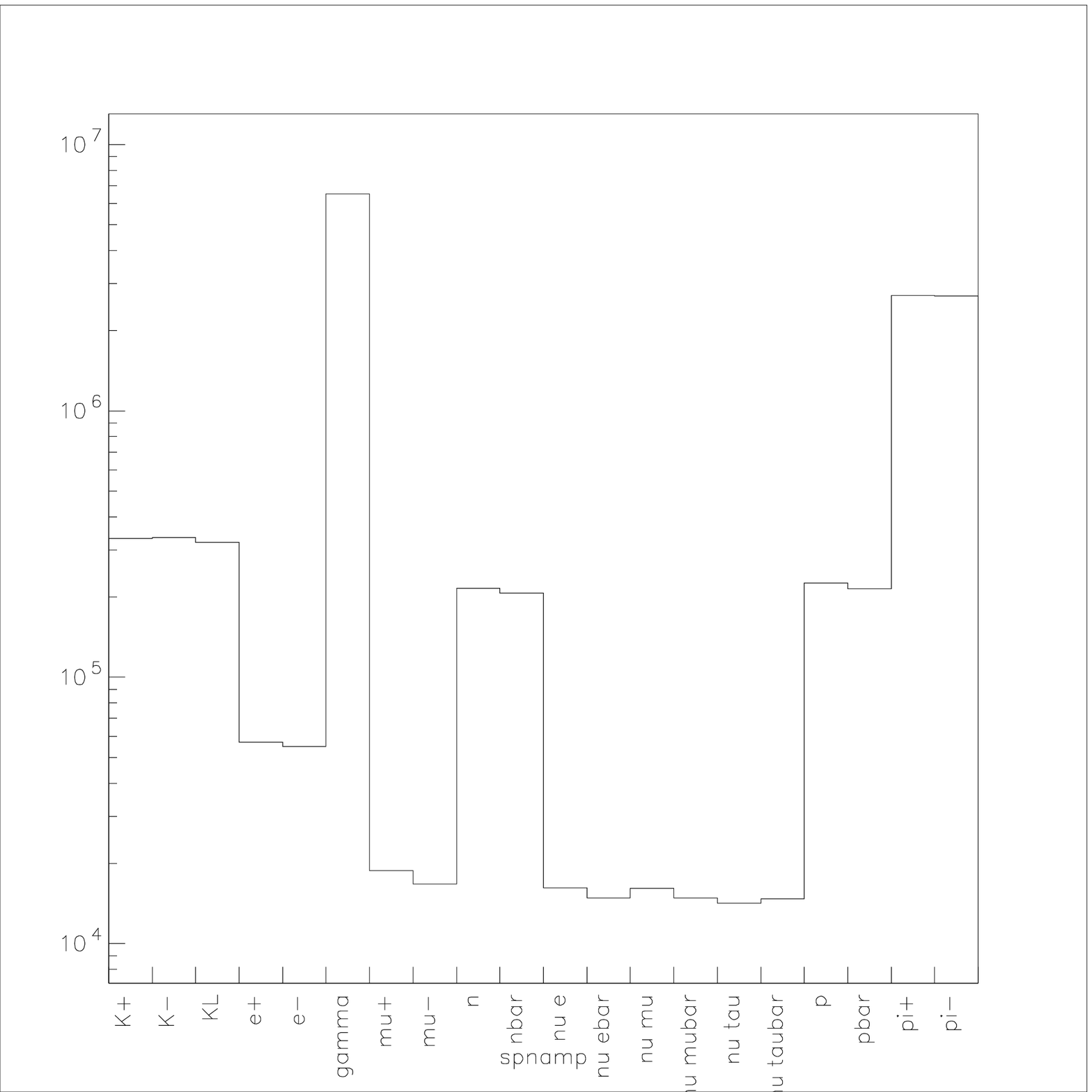}
}
\caption{Final abundance of Standard Model particles produced during the
evaporation of $10\,$TeV black holes for a run of 10000 events.
Upper graphs:
standard CHARYBDIS generator with $M_{\rm f} =1\,$TeV
(left panel) and $M_{\rm f} =1\,$GeV (right panel).
Lower graphs:
modified code with the temperature $T_1$ (left panel)
and $T_2$ (right panel).
\label{part} }
\end{figure}
\begin{figure}[ht]
\centering{
\includegraphics[width=0.45\textwidth]{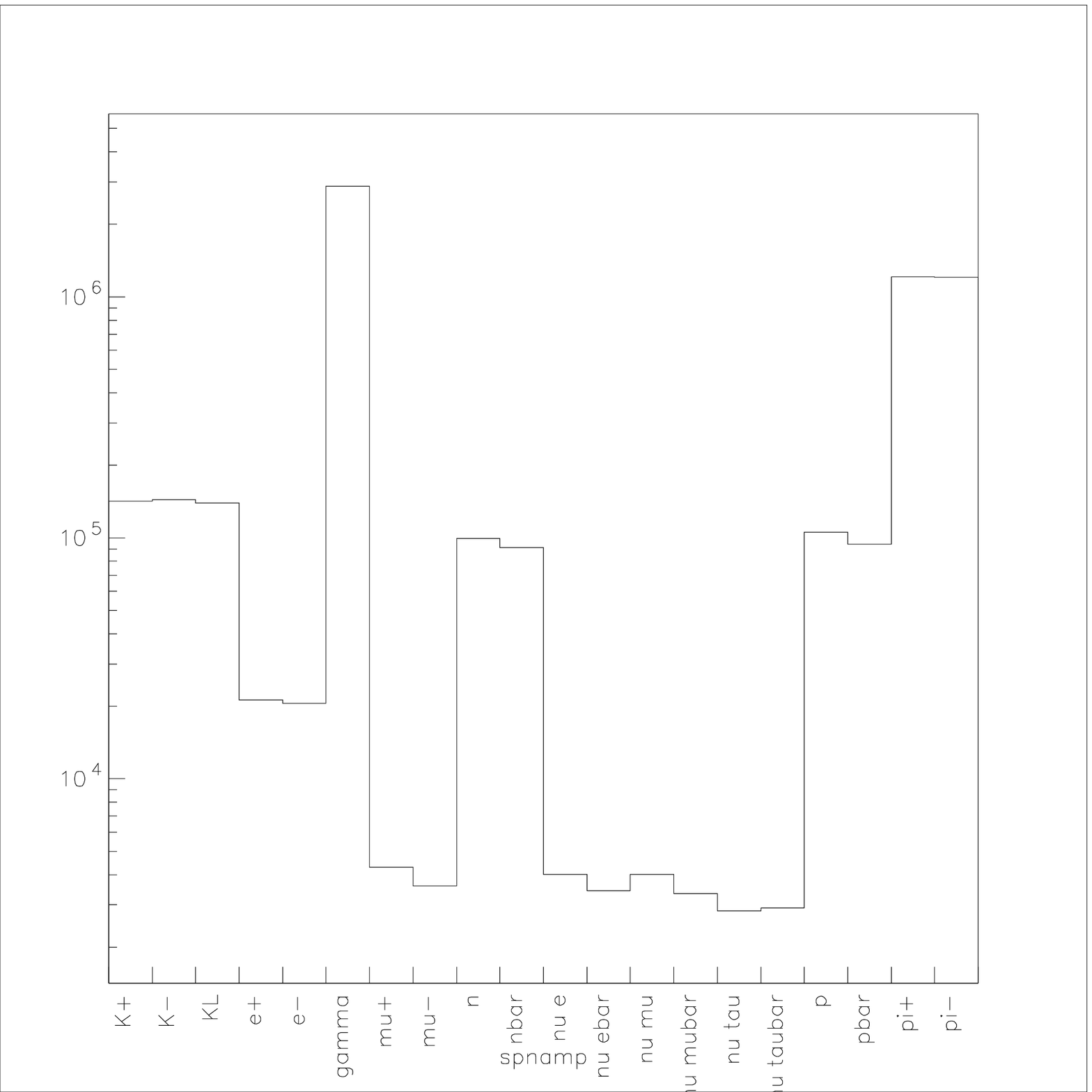}
\hspace{0.5cm}
\includegraphics[width=0.45\textwidth]{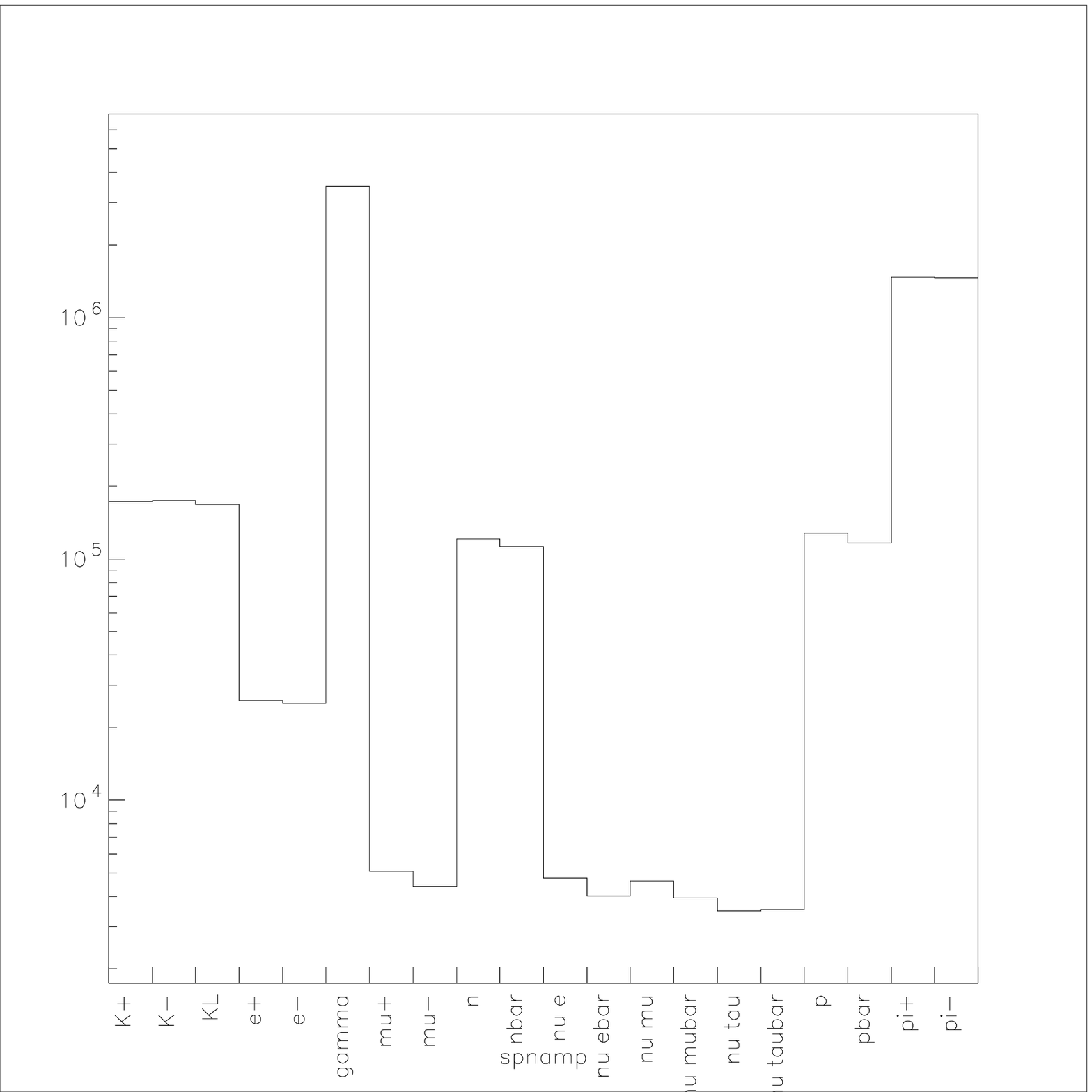}
} \vspace{0.5cm}
\\
\centering{
\includegraphics[width=0.45\textwidth]{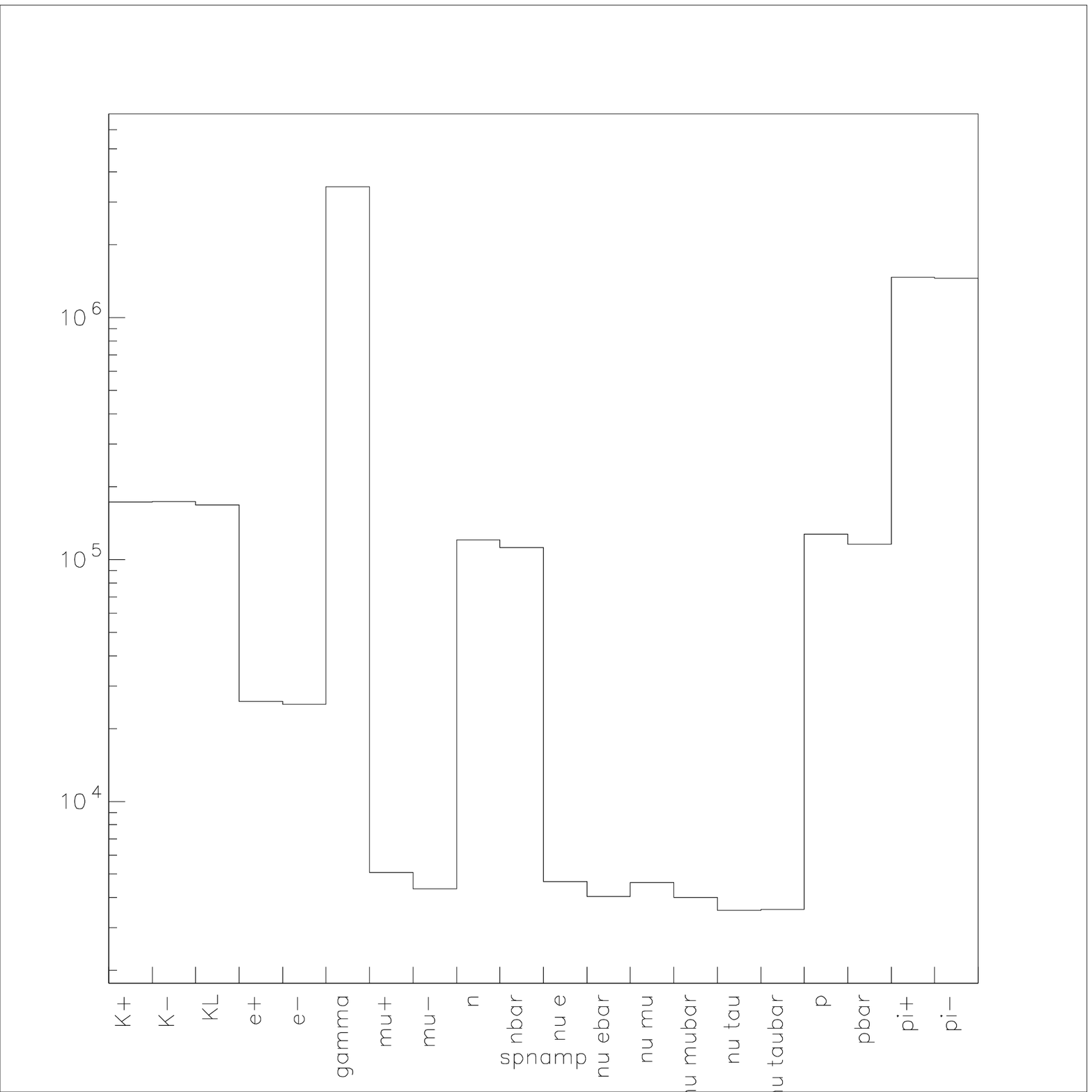}
\hspace{0.5cm}
\includegraphics[width=0.45\textwidth]{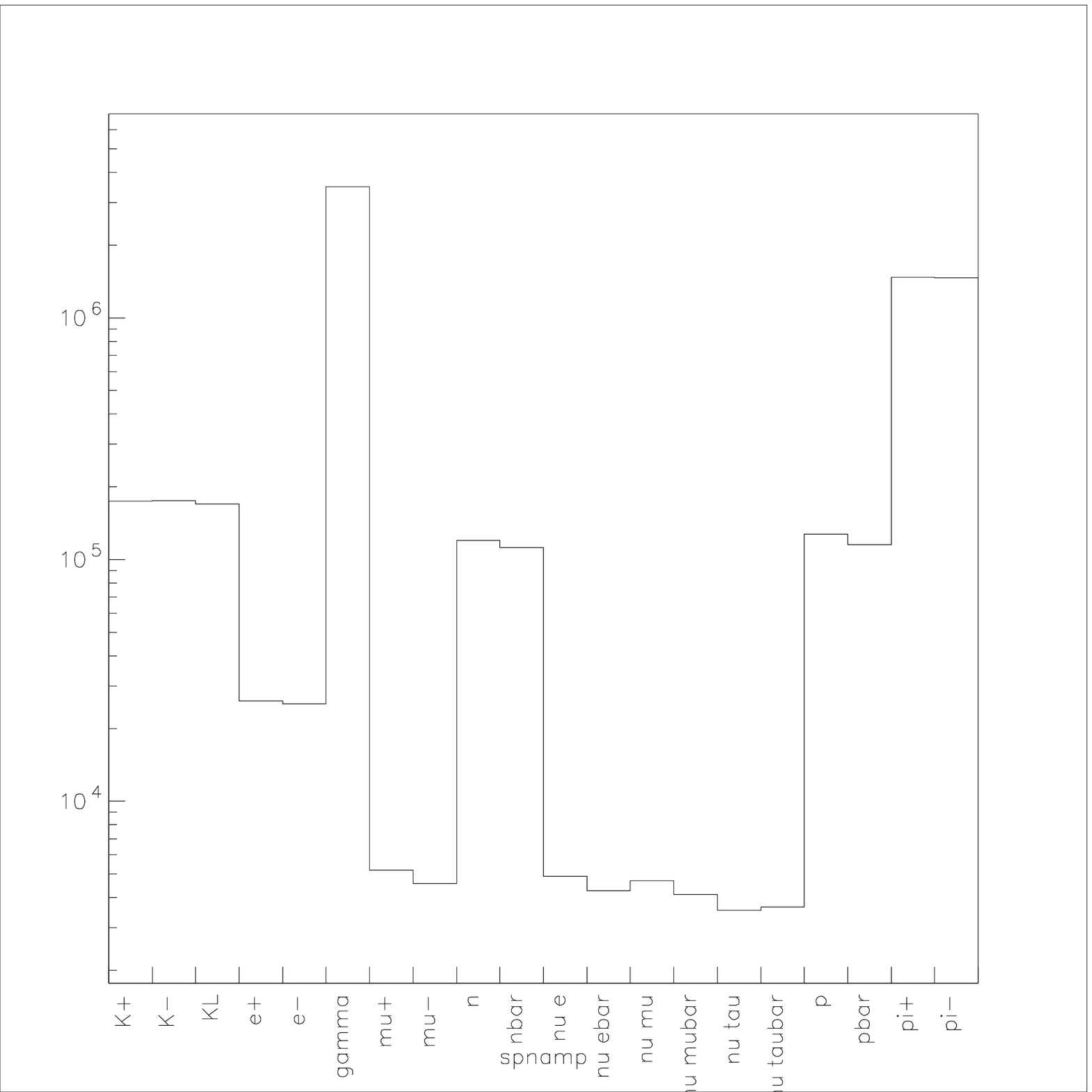}
}
\caption{Final abundance of Standard Model particles produced during the
evaporation of $2\,$TeV black holes for a run of 10000 events.
Upper graphs:
standard CHARYBDIS generator with $M_{\rm f} =1\,$GeV
Lower graphs:
modified code with the temperature $T_1$ (left panel)
and $T_2$ (right panel).
\label{part2tev} }
\end{figure}
\begin{figure}[ht]
\centering{
\includegraphics[width=0.45\textwidth]{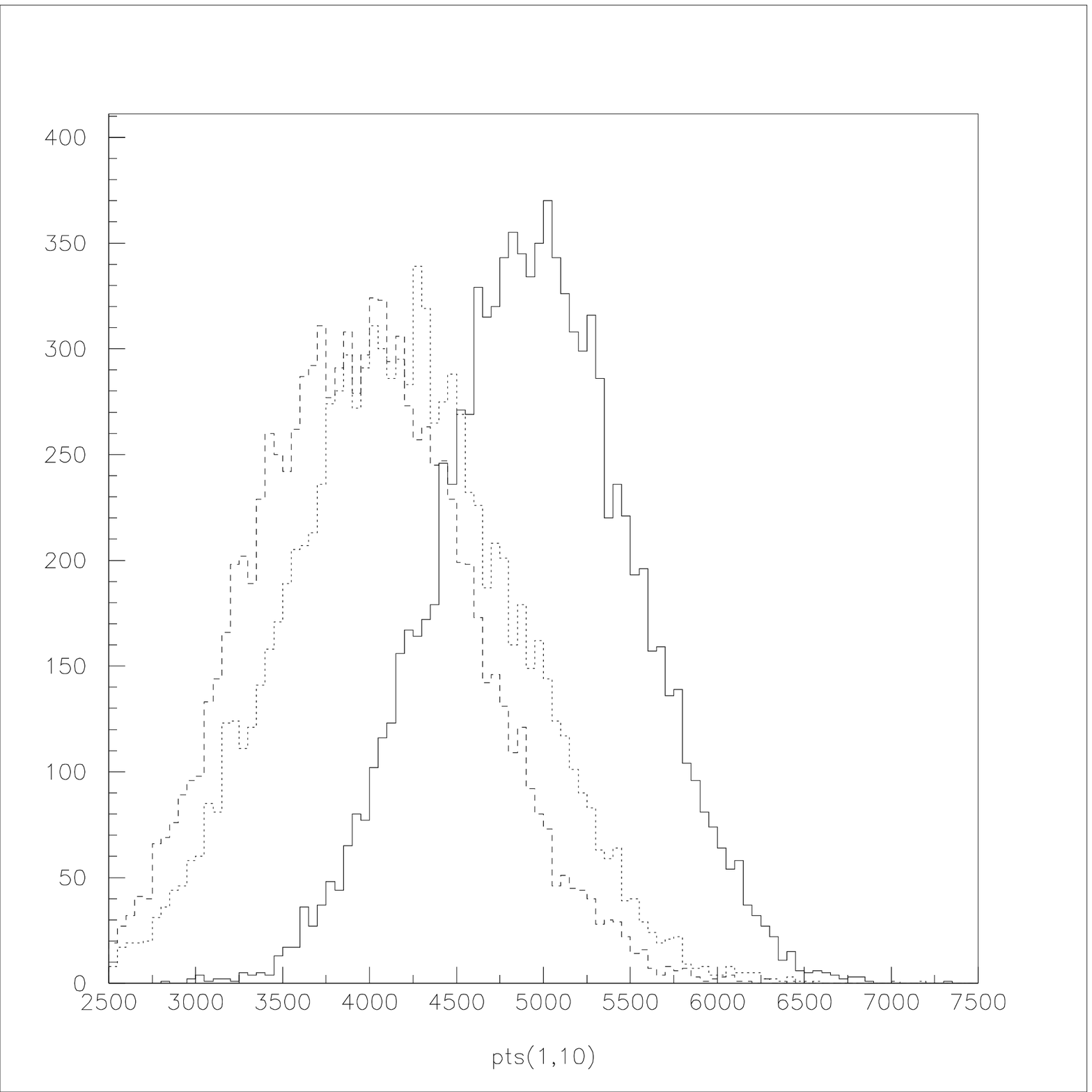}
\hspace{0.5cm}
\includegraphics[width=0.45\textwidth]{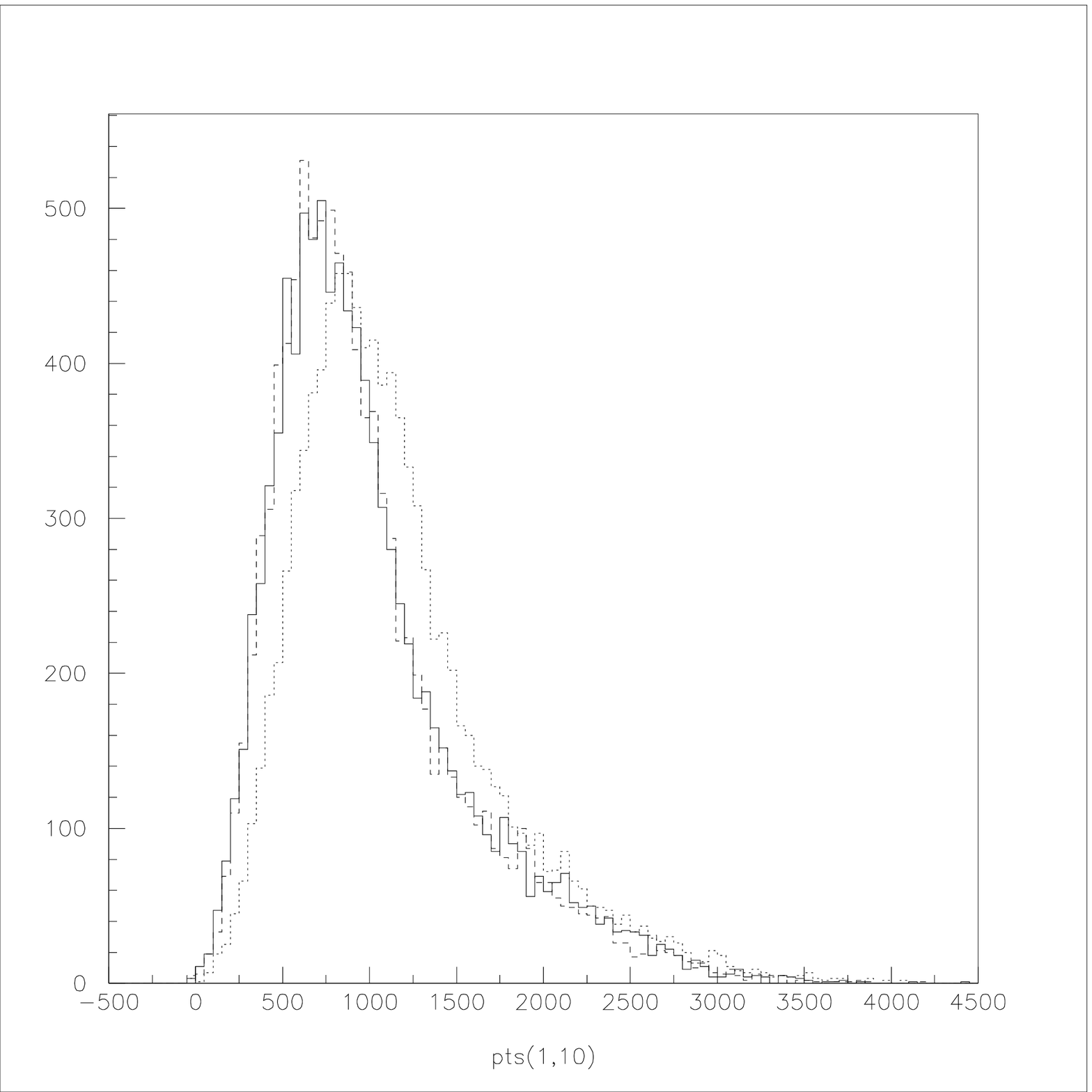}}
\caption{Statistical distribution of the sum of particle
transverse momenta including only particles with
transverse momentum higher than $30\,$GeV for 10000 events
for the decay of $10\,$TeV black holes (left graph)
and of $2\,$TeV black holes (right graph):
standard CHARYBDIS generator with $M_{\rm f} =1\,$TeV
(pointed line) and $M_{\rm f} =1\,$GeV (dashed line),
modified code with the temperature $T_1$ (solid line)
and $T_2$ (indistinguishable from dashed line).
\label{pts1-10} }
\end{figure}
\begin{figure}[ht]
\centering{
\includegraphics[width=0.45\textwidth]{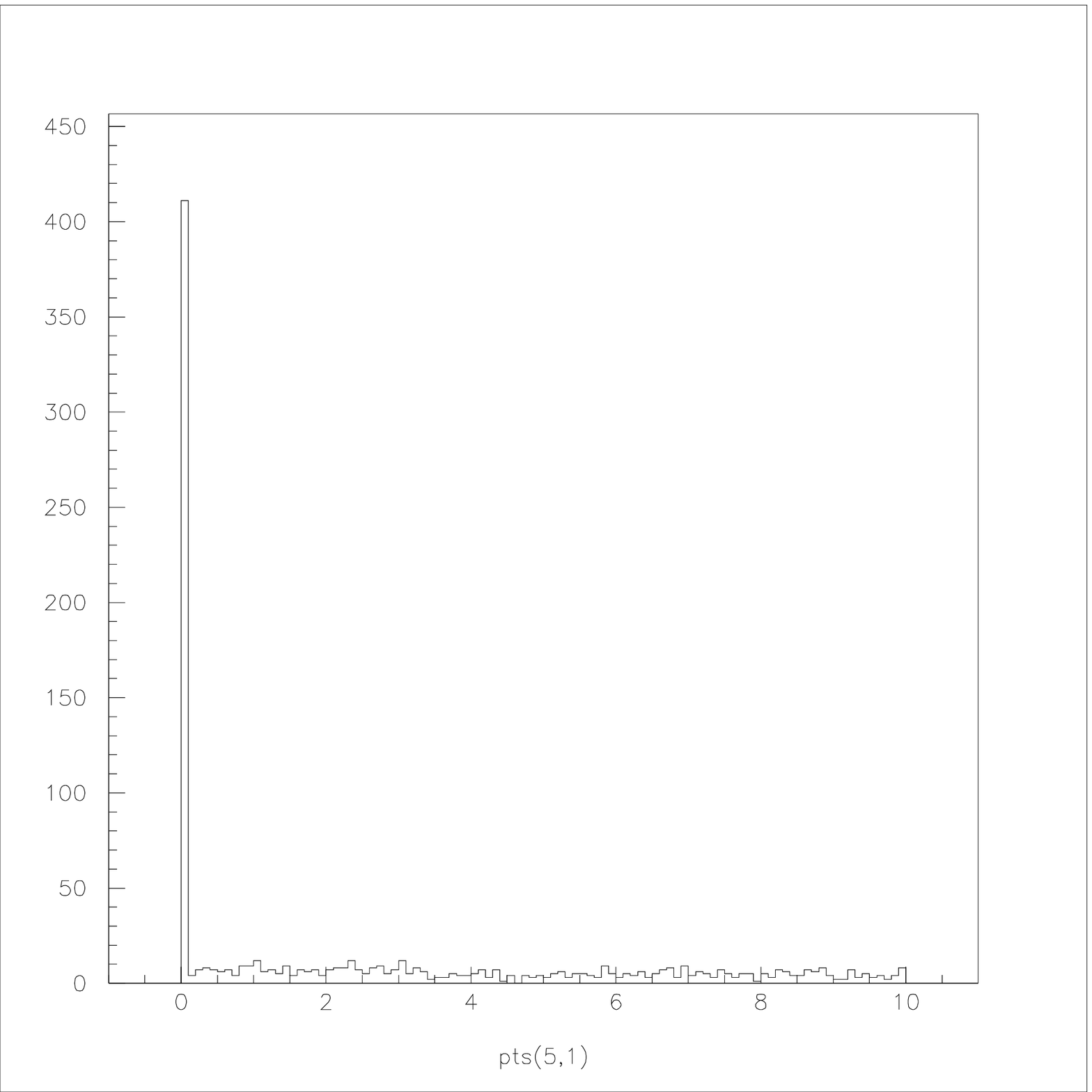}
\hspace{0.5cm}
\includegraphics[width=0.45\textwidth]{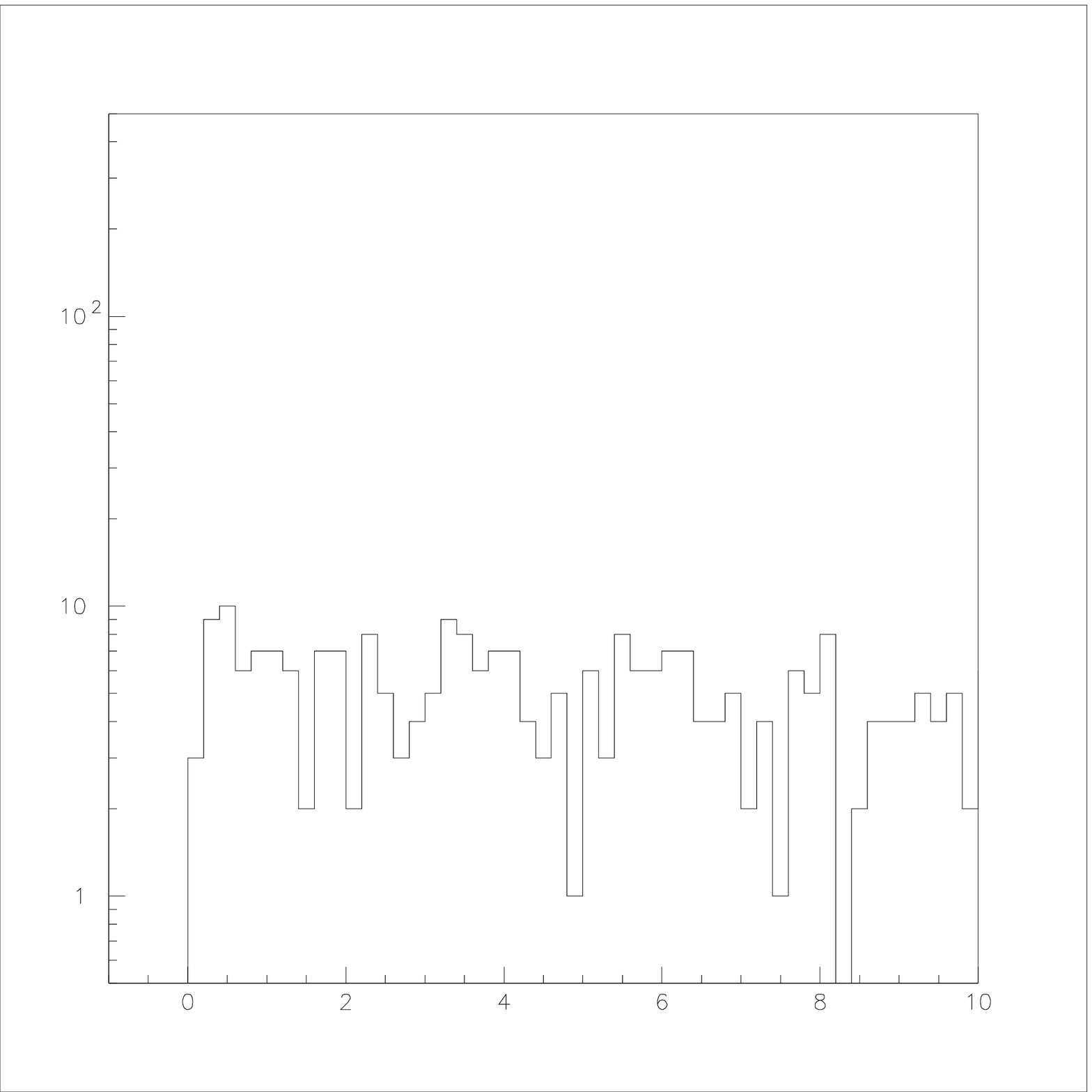}
}
\caption{Sum of the muon transverse momenta in the range
$0-10\,$GeV including only particles with
transverse momentum higher than $30\,$GeV for the decay
of $10\,$TeV black holes.
Left panel:
standard CHARYBDIS generator with $M_{\rm f} =1\,$TeV.
Right panel:
modified code with the temperature $T_1$ for $M_{\rm f} =1\,$GeV.
\label{pts5-1b} }
\end{figure}
The primary output includes quarks and gluons that cannot be
detected.
We therefore used Pythia~\cite{pythia} to model the parton evolution
and hadronization of quarks and gluons emitted during the evaporation
(or not participating to the black hole events).
This produced the final aspect of the events as would be seen
by a detector at the LHC.
\par
In Fig.~\ref{part}, we show the final particle abundance for the standard
Hawking temperature and $M_{\rm f}=1\,$TeV or $M_{\rm f}=1\,$GeV
and the modified temperatures $T_1$ and $T_2$ with $M_{\rm f}=1\,$GeV.
The first thing one notes is that the standard Hawking temperature
for both $M_{\rm f}=1\,$TeV or $M_{\rm f}=1\,$GeV and the temperature
$T_2$ now produce very similar numbers of particles.
In contrast to the primary output, the temperature $T_1$ now
produces a number of photons and pions smaller than the other cases.
The number of muons and electrons in the final state is however
still higher (by a factor around 2) for $T_1$ with respect to all other cases.
In the case of a black hole with initial mass of $2\,$TeV,
Fig.~\ref{part2tev} shows no evident difference in the number of produced
particles for the four cases previously discussed.
\par
Another remarking feature of the events produced with the temperature
$T_1$ is that the distribution of the sum of transverse momenta for particles with
transverse momentum higher than $30\,$GeV is peaked around a
larger value (about $5\,$TeV) compared to the results for the other 
three cases, as shown in the left graph of Fig.~\ref{pts1-10}
(in the right graph, the analogue distributions for $M_0=2\,$TeV are showed).
If one counts all particles, however, such distributions
show no appreciable difference.
Finally, we remark that the distribution for the sum of the muon transverse
momenta for the temperature $T_1$ does not show a peak around zero,
contrary to all other cases.
The same feature appears if one only counts muons with transverse momentum
over $30\,$GeV see Fig.~\ref{pts5-1b}.
\section{Conclusions}
\setcounter{equation}{0}
We have considered possible modifications to the standard picture
of the decay of micro-black holes that might be produce at the LHC.
Inspired by the microcanonical description of the Hawking evaporation
and other theoretical approaches in the literature, we have studied modified
statistical laws for the emitted particles described by different
mass-dependent temperatures $T_1$ and $T_2$ which have a regular
behavior for vanishing black hole mass.
We have also required that energy be conserved all the way to the end of the
evaporation at a scale $M_{\rm f}\ll M_{\rm G}$.
We have not explicitly considered the case in which
black holes leave stable remnants, since that has already been
reported in Ref.~\cite{Koch:2005ks}.
Further, our temperature
$T_2$ corresponds to a zero-temperature remnant of finite mass which suddenly
decays in a few particles.
\par
The numerical simulations we have run so far suggest that this kind of
modifications, if present, might have significant and testable
effects for black holes produced with a (relatively) large initial mass
(of the order of ten times the fundamental scale of gravity $M_{\rm G}$).
On assuming that $M_{\rm G}=1\,$TeV, the LHC at full luminosity is expected to
produce around three $10\,$TeV black holes per day~\cite{Giddings3}.
Such a production rate, although smaller than the $1\,$Hz
for $5\,$TeV black holes (and even larger for $2\,$TeV black holes)
is however large enough to collect sufficient data to compare with
the results we have presented here and their signature should also be easier
to identify than that for lighter black holes.
For example, one expects to see differences in the number of particles
produced and in their total transverse momentum.
It is therefore important to take into account these possibilities in
any detailed simulation of the evaporation process whose outputs
will have to be confronted with forthcoming data.
\section*{Acknowledgements}
R.~C. thanks M.~Cavaglia, S.~Hsu,  I.~Lazzizzera and the CMS group in Trento,
T.G.~Rizzo and collaborators at SLAC.
We would like to thank H.~Menghetti for helpful discussions.
We remain indebted with V.~Vagnoni and the LHCb group in Bologna for the
early stages of the present work. 
\end{document}